# Time-Optimal Directed $q$-Analysis

**Felix Windisch** ✉ ⓘ
Graz University of Technology, Austria

**Florian Unger** ✉ ⓘ
Graz University of Technology, Austria

## Abstract

Directed $q$-analysis is a recent extension of $q$-analysis, an established method for extracting structure from networks, to directed graphs. Until recently, a lack of efficient algorithms heavily restricted the application of this technique: Previous approaches scale with the square of the input size, which is also the maximal size of the output, rendering such approaches worst-case optimal. In practice, output sizes of relevant networks are usually far from the worst case, a fact that could be exploited by an (efficient) output-sensitive algorithm.

We develop such an algorithm and formally describe it in detail. The key insight, obtained by carefully studying various approaches to directed $q$-analysis and how they relate to each other, is that inverting the order of computation leads to significant complexity gains. Targeted precomputation and caching tactics further reduce the introduced overhead, enough to achieve (under mild assumptions) a time complexity that is linear in output size. The resulting algorithm for performing directed $q$-analysis is shown to be time-optimal.

**2012 ACM Subject Classification** Theory of computation → Graph algorithms analysis

**Digital Object Identifier** 10.4230/LIPIcs.Preprint..1

**Acknowledgements** We would like to express our gratitude for their time, ideas and general help (alphabetic order): Olga Diamanti, Bianca Dornelas, Michael Kerber, Robert Legenstein, Henri Riihimäki, Markus Steinberger.

## 1 Introduction

Classical $q$-analysis is an established tool for extracting useful information about the structure of graphs [1, 2, 3]. Originally developed for undirected graphs, it was recently generalized to directed graphs [8], in what is known as *directed q-analysis*.

Our starting point is the recently published work [11], in which techniques from directed $q$-analysis were successfully applied to analyse connectomic graphs, i.e. the directed graphs that represent neuron-synapse-level connectivity in brains. So far processing these connectomes has been notoriously difficult or downright impossible, due to the prohibitive computational cost. The authors of [11] gave improvements to various key elements of directed $q$-analysis, most notably in the very definition itself, but maybe even more impactful for future research released a software library boasting speed-ups in the millions compared to previous implementations.

The focus of [11] was on application, but there was no analysis, or even detailed description of the algorithms that compute directed $q$-analysis. From a more theoretical standpoint, this leaves major gaps. In particular, we set out to address the following questions:

- How do the algorithms utilized in [11] (and also [8]) work?
- Are they correct?
- What is their asymptotic time complexity?

## 1.1   A Crash Course on Directed $q$-Analysis

We start out with a very brief, high-level outline of directed $q$-analysis, to better understand the associated computational challenges. We keep the discussion here as abstract as possible to follow the key insights. A more application-oriented introduction to directed $q$-analysis, providing intuition and details about its practical relevance may be found in [11], see further [8] or [10] for different approaches to introduce the matter.

**What is $q$-Analysis?**   In a nutshell, directed $q$-analysis transforms a directed graph $G$ into another, typically much larger, but sparser, directed graph $\mathcal{Q}$. This new graph $\mathcal{Q}$ features as vertices all the *simplices* (acyclic and dense subgraphs) of $G$. Edges in the $\mathcal{Q}$-graph then represent "interaction" between simplices: for given integer parameters $q, i, j$, presence of an edge in $\mathcal{Q}$ indicates that the two incident simplices share a $q$-sized subsimplex in a way that matches the *directions $i, j$* (see Section 2 for a formal description) — we then call the simplices *$(q, i, j)$-near*. While fast enumeration of all simplices is a solved problem (see e.g. [4, 5]), the novel computational challenge lies in finding all the *edges* of $\mathcal{Q}$.

Note that there are currently two subtly different definitions of $(q, i, j)$-nearness: The more classic definition introduced in [8], which we refer to as $\widehat{(q, i, j)}$-nearness, and a more novel definition introduced in [11], to which we refer as $(q, i, j)$-nearness.

**Original Approach: Top-Down.**   The approach utilized in the first exploratory work on directed $q$-analysis, [8], exploited the fact that checking two simplices for $\widehat{(q, i, j)}$-nearness (respectively $(q, i, j)$-nearness) is comparatively straight-forward. A straightforward extension to find *all* the edges of $\mathcal{Q}$ is then to simply check all pairs of simplices, leading to an $\Omega(n^2)$-algorithm in the number of simplices. We refer to this as Top-Down approach, since one starts at the top (large simplices) and works one's way towards the bottom (small simplices). Unfortunately, this algorithm, which was the state of the art prior to [11], does not scale well to graphs with millions of simplices and only sparse interaction between them. But these are precisely the type of graphs commonly investigated in e.g. Computational Neuroscience. A common example is the Blue Brain Project's partial stochastic reconstruction of a rodent somatosensory cortex [7], henceforth referred to as BBP. For $q = 4$, BBP contains 810K simplices but only 818K edges in the $\mathcal{Q}$-graph. This implies that around 99.9999% of all checks necessary for the Top-Down approach discover no edge and are thus — in retrospect — superfluous.

**An Output-Sensitive Approach: Bottom-Up.**   For sparse results like with the BBP, an "output-sensitive" algorithm would scale much better. It would require computation if and only if there is an actual interaction between two simplices to be found. The key insight we describe here is that, instead of examining each individual pair of simplices and looking for shared subsimplices, one can instead look at the smallest simplices where interaction could happen and then propagate those results "upward" to larger simplices that contain them. This approach, without further refinement, is referred to as the Bottom-Up approach in this paper.

**The Time-Optimal Hybrid Approach.**   While the Bottom-Up approach features an output-sensitive runtime and is thus, in practice, already much faster than the Top-Down approach, it still suffers from the computationally costly propagation of simplicial interaction to higher levels. To mitigate this, we describe, in detail, the so-called Hybrid approach, which in a single top-down pass precomputes and caches certain information to drastically speed up the (previously prohibitively expensive) upward propagation. Indeed this approach reduces, under mild conditions, the computational effort per edge in the $\mathcal{Q}$-Graph to $\mathcal{O}(1)$ and thereby allows development of an asymptotically (w.r.t. graph size) time-optimal algorithm to calculate the directed $q$-connectivity graph $\mathcal{Q}$.

## 1.2 Contributions

The main contributions of this paper are:
- a formal exploration of the relation between $\widehat{(q,i,j)}$-nearness and $(q,i,j)$-nearness,
- a formal presentation of the Hybrid algorithm,
- a proof of the time-optimality of the Hybrid approach.

We further formalize the Top-Down algorithm, allowing for easy analysis of correctness and time complexity. An equally thorough formalization of the Hybrid approach eases analysis of correctness for it as well. We report on parallelization efforts for the Hybrid and Bottom-Up approach. We reason that in some scenarios the Bottom-Up approach would be superior one.

## 1.3 Outline

Section 2 introduces all necessary definitions and provides details and proofs to the relation between the two concurrent definitions of directed $q$-analysis.
Section 3 formalizes the Top-Down approach and Hybrid approaches.
Section 4 analyses the asymptotic runtime of all presented algorithms, in particular the time optimality of the Hybrid approach.
Section 5 summarizes different approaches to parallelize the Hybrid algorithm and briefly introduces the Bottom-Up algorithm.
Section 6 discusses limitations of the Hybrid approach and potential solutions.

## 2 Preliminaries

This section deals with the purely mathematical aspects of directed $q$-analysis. We list all definitions necessary for a formal understanding of directed $q$-analysis, then the two competing definition currently found in the literature [8, 11] and conclude with a formal analysis of their relations. While not every lemma and proposition is strictly necessary to understand the development of the algorithm, following them and their proofs both deepens understanding of their relation and helps with understanding the algorithms in the next section.

### 2.1 Graphs and Simplices

▶ **Definition 1** (Graphs).
- *A simple directed graph is a pair $G = (V, E)$ of a finite set of vertices $V$ and a relation $E \subseteq (V \times V) \setminus \Delta_V$, where $\Delta_V = \{(v,v) \mid v \in V\}$.*
- *The set of successors of a vertex $v$ in $G = (V, E)$ is denoted as $\delta_G^+(v) := \{w \in V \mid (v,w) \in E\}$ and analogously the predecessors are denoted as $\delta_G^-(v) := \{w \in V \mid (w,v) \in E\}$.*
- *Let $G = (V, E)$ and $S = (V_S, E_S)$ be a graph with $V_S \subseteq V$ and $E_S \subseteq E$. Then $S$ is called a subgraph of $G$.*

Since we will only use simple directed graphs, we will henceforth refer to them as *graphs*.

▶ **Definition 2** (Cliques and Simplices). *Let $G = (V, E)$ be a directed graph.*

- A $d$-*simplex is a totally ordered subset* $(v_0 v_1 \ldots v_d)$ *of $V$ such that additionally* $(v_i, v_j) \in E$ *for all $i < j$. They are denoted by* $(v_0 v_1 \cdots v_d)$.
- The dimension *of a simplex $\sigma$, denoted $\dim(\sigma)$ is equal to $|\sigma| - 1$.*
- *We call a simplex $\mu_\sigma$ a* face *of $\sigma$ if its vertices are a subset of the vertices of $\sigma$ and the vertices occur in the same order. We write $\mu_\sigma \hookrightarrow \sigma$ or $\sigma$ includes $\mu_\sigma$.*

▶ **Definition 3** (Directed Flag Complexes and Face Maps).
- *Let $G = (V, E)$ be a simple directed graph and $D$ the dimension of the highest-dimensional simplex in $G$. The directed flag complex $\Sigma$ is a tuple of length $D+1$ storing all the $d$-dimensional simplices in its respective fields, i.e. $\Sigma := (\Sigma_0, \Sigma_1, \ldots, \Sigma_D)$, in particular $\Sigma_0 = V$ and $\Sigma_1 = E$.*
  *We write $\Sigma_{>q}$ for the subset of simplices with dimension greater than $q$.*
- *Let $\Sigma_n \ni \sigma = (\sigma_0 \ldots \sigma_n)$. The face maps*

$$\hat{d}_i(\sigma) := \begin{cases} \sigma \setminus (\sigma_i) & \text{if } i \leq n \\ \sigma \setminus (\sigma_n) & \text{if } i > n \end{cases}$$

  *associate the simplex with its $i$th face.*
  *Whenever $i \leq n$ can be guaranteed, we omit the hat and use $d_i(\sigma) := \sigma \setminus (\sigma_i)$. We further write $\hat{d}_\infty$ to refer to removing the last index when not explicity known.*
- *Let $\sigma \in \Sigma_p$. The* coface *operator $\mathrm{coF}_i(\sigma) = \{\tau \in \Sigma_{p+1} \mid d_i(\tau) = \sigma\}$ is the left inverse of $d_i$. Note that $\mathrm{coF}_i$ yields a set of simplices and thus is only a one-sided inverse of $d_i$.*
- *Let $\mathrm{faces}(\sigma, n) := \{\tau \in \Sigma_n \mid \tau \hookrightarrow \sigma\}$ denote the set of all $n$-dimensional faces of $\sigma$.*

See Figure 1 as an example of a graph and its flag complex.

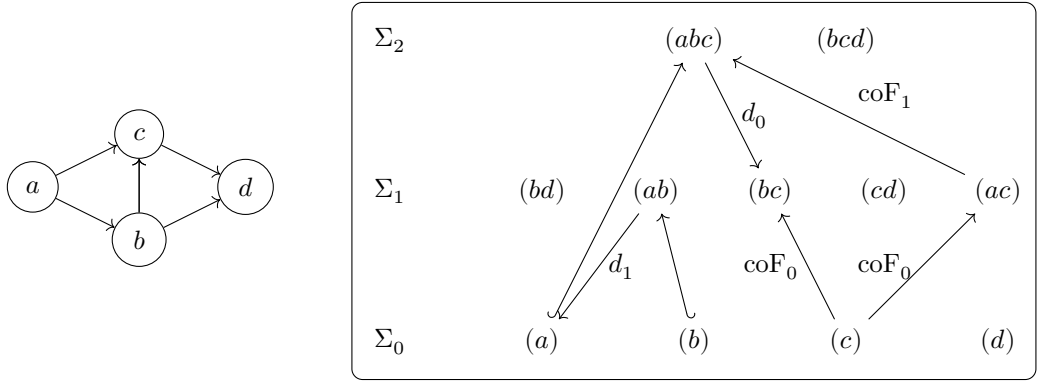

**Figure 1** Directed graph and corresponding flag complex with selected face ($d_i$), coface ($\mathrm{coF}_i$) and inclusion ($\hookrightarrow$) relations marked.

## 2.2 Directed $q$-Analysis

In [8] the concept of $q$-analysis was extended to directed graphs with the following definition:

▶ **Definition 4** (original directed $q$-nearness). *Let $\Sigma$ be a directed flag complex and $(\sigma, \tau)$ be an ordered pair of simplices $\sigma, \tau \in \Sigma_{>q}$. Let $(\hat{d}_i, \hat{d}_j)$ be an ordered pair of face maps. Then $\sigma$ is $\widehat{(q, i, j)}$-near to $\tau$ if either of the following conditions is true:*

  **[I]** $\sigma \hookrightarrow \tau$,
  **[II]** $\hat{d}_i(\sigma) \hookleftarrow \alpha \hookrightarrow \hat{d}_j(\tau)$ *for some $\alpha \in \Sigma_q$.*

More recently, [11] introduced a modified definition, claiming better interpretability of these results as their motivation:

▶ **Definition 5** (novel directed q-nearness). *Let $\Sigma$ be a directed flag complex and $(\sigma, \tau)$ be an ordered pair of simplices $\sigma, \tau \in \Sigma_{>q}$. Let $(d_i, d_j)$ be an ordered pair of face maps with $i, j \in \{0, ..., q+1\}$. Then $\sigma$ is $(q, i, j)$-near to $\tau$ if either of the following conditions is true:*

[I] $\sigma \hookrightarrow \tau$,
[II] *There exist a q-simplex $\alpha \in \Sigma_q$ and two $(q+1)$-simplices $\mu_\sigma \hookrightarrow \sigma, \mu_\tau \hookrightarrow \tau$ such that $d_i(\mu_\sigma) = \alpha = d_j(\sigma_\tau)$.*

We cite Figure 2 from [11], as it gives a schematic overview of criterion [II] for both the new and original definition on directed graphs.

In accordance to [11], we refer to simplices as being $(q, i, j)$-near respectively $\widehat{(q, i, j)}$-near by criterion [I] (inclusion) or criterion [II] (shared face) for both definitions. Hooked arrows ($\sigma \hookrightarrow \tau$) denote q-nearness by criterion [I], while dashed arrows ($\sigma \dashrightarrow \tau$) denote q-nearness by criterion [II].

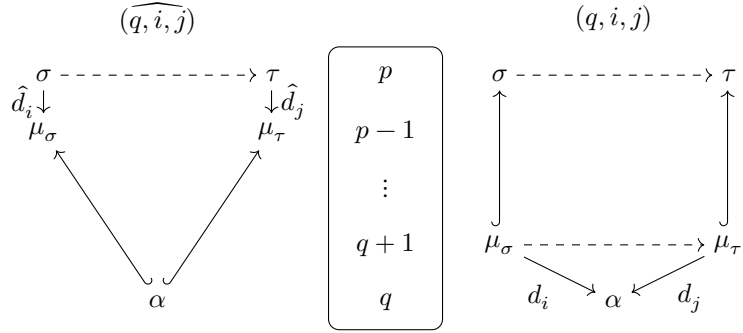

**Figure 2** Condition for two simplices $\sigma, \tau$ to be $\widehat{(q, i, j)}$-near (left) or $(q, i, j)$-near (right) by criterion [II]. Dimension indicator in the middle.

▶ **Definition 6** ($(q, i, j)$-Digraphs). *Let $G$ be a simple, directed graph with directed flag complex $\Sigma$. The $(q, i, j)$-digraph $\mathcal{Q} = \{\Sigma, E^{\mathcal{Q}}\}$ of $G$ contains an edge $(\sigma, \tau)$ if $\sigma \in \Sigma$ is $(q, i, j)$-near to $\tau \in \Sigma$. Analogously, $\hat{\mathcal{Q}} = \{\Sigma, E^{\hat{\mathcal{Q}}}\}$ contains connections between $\widehat{(q, i, j)}$-near simplices.*

When referring to particular q-digraphs, they are referred to as the $(q, i, j)$- or $\widehat{(q, i, j)}$-digraph of $G$.

## 2.3 Comparing $(q, i, j)$-nearness and $\widehat{(q, i, j)}$-nearness

We explore differences and similarities between $(q, i, j)$-nearness and $\widehat{(q, i, j)}$-nearness. While, with the exception of 10, none are strictly required to understand Section 3, they help with understanding these rather abstract concepts and the subtle differences between both competing notions of directed q-analysis.

### 2.3.1 Similarities

The following three results describe similarities between both definitions and were (without proof) already stated in [11].

▶ **Lemma 7.** *If $\sigma, \tau \in \Sigma_{q+1}$ are $(q, i, j)$-near, they are also $\widehat{(q, i, j)}$-near and vice versa.*

**Proof.** Due to the reflexivity of inclusions ($\alpha \hookrightarrow \alpha$), the definitions of $\widehat{(q,i,j)}$-nearness and $(q,i,j)$-nearness in the case of $(q+1)$-dimensional simplices are identical:

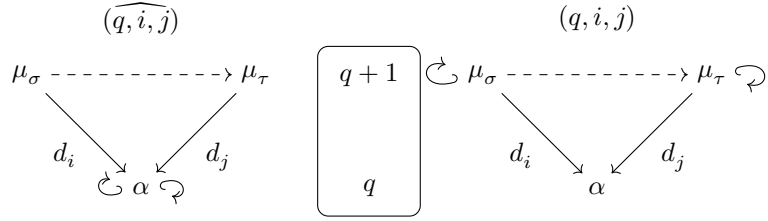

◀

The definitions also coincide for particularly important pairs (see [11]) of $i$ and $j$:

▶ **Proposition 8.** *Let $G$ be a directed, simple graph with associated $(q,0,\infty)$-digraph $\mathcal{Q}$ and $\widehat{(q,0,\infty)}$-digraph $\hat{\mathcal{Q}}$. Then, $\mathcal{Q} = \hat{\mathcal{Q}}$.*
*This also holds true for the directions $(q,0,0)$, $(q,\infty,0)$ and $(q,\infty,\infty)$.*

**Proof.** All edges between simplices that fulfill criterion [I] are trivially contained in both $\mathcal{Q}$ and $\hat{\mathcal{Q}}$. It is sufficient to proof the statement for simplices that are near by criterion [II]:

$\hat{\mathcal{Q}} \subseteq \mathcal{Q}$: Assume two simplices $\sigma = (\sigma_0 \sigma_1 \cdots \sigma_n)$ and $\tau = (\tau_0 \tau_1 \cdots \tau_m)$ are $\widehat{(q,0,\infty)}$-near. Then by definition there exists an $\alpha \in \Sigma_q$ such that $\hat{d}_0(\sigma) \hookleftarrow \alpha \hookrightarrow \hat{d}_\infty(\tau)$. Consider the faces $\mu_\sigma = (\sigma_0 \alpha_0 \alpha_1 \cdots \alpha_q) \hookrightarrow \sigma$ and $\mu_\tau = (\alpha_0 \alpha_1 \cdots \alpha_q \tau_m) \hookrightarrow \tau$, which share the $(q)$-face $\alpha$. From $\sigma \hookleftarrow d_0(\mu_\sigma) = \alpha = d_\infty(\mu_\tau) \hookrightarrow \tau$, it follows that $\sigma$ and $\tau$ are $(q,0,\infty)$-near. We note that $\mu_\sigma$ and $\mu_\tau$ are well defined and do not contain duplicate vertices ($\sigma_0 \not\hookrightarrow \alpha \not\hookleftarrow \tau_m$), because $\alpha \hookrightarrow \hat{d}_0(\sigma) = (\sigma_1, \cdots \sigma_n)$ and $\alpha \hookrightarrow \hat{d}_\infty(\tau) = (\tau_0 \cdots \tau_{n-1})$.

$\mathcal{Q} \subseteq \hat{\mathcal{Q}}$: Consider two $(q,0,\infty)$-near simplices $\sigma = (\sigma_0 \sigma_1 \cdots \sigma_n)$ and $\tau = (\tau_0 \tau_1 \cdots \tau_m)$. By definition, there are two faces $\mu_\sigma \hookrightarrow \sigma$ and $\mu_\tau \hookrightarrow \tau$, such that $d_0(\mu_\sigma) = d_\infty(\mu_\sigma) = \alpha$. Note that $\sigma_0 \notin d_0(\mu_\sigma)$ and $\tau_m \notin d_\infty(\mu_\tau)$, meaning neither can be contained in $\alpha$. From this and $\sigma \hookleftarrow \alpha \hookrightarrow \tau$, we can conclude that $d_0(\sigma) \hookleftarrow \alpha \hookrightarrow d_\infty(\tau)$ and that $\sigma, \tau$ are $\widehat{(q,0,\infty)}$-near

The proof for the other directions is analogous with different indices. ◀

▶ **Proposition 9.** *If two simplices $\sigma$ and $\tau$ are $(q,i,j)$-near, they are also $\widehat{(q,k,l)}$-near for some $k,l \in \mathbb{N}$ and vice versa.*

**Proof.** Should $\sigma \hookrightarrow \tau$ or $\sigma \hookleftarrow \tau$ hold true, both definitions trivially coincide for any $i$ and $j$. Otherwise, both $(\sigma \backslash \tau) \neq \emptyset$ and $(\tau \backslash \sigma) \neq \emptyset$ hold and they must share a $(q)$-face $\alpha$.

$\Rightarrow$: Consider the case where $\sigma$ is $(q,i,j)$-near to $\tau$. Let $k$ be the index of any vertex $\sigma_k \in (\sigma \backslash \tau)$ in $\sigma$ and $l$ the index of any $\tau_l \in (\tau \backslash \sigma)$ in $\tau$. From $\alpha \hookrightarrow \sigma$ and $\sigma_k \notin \alpha$ follows that $\alpha \hookrightarrow d_k(\sigma)$. Similarly, $\alpha \hookrightarrow \tau$ and $\tau_l \notin \alpha$ imply $\alpha \hookrightarrow d_l(\tau)$. The definition of $\widehat{(q,k,l)}$-near is fulfilled, since $d_k(\sigma) \hookleftarrow \alpha \hookrightarrow d_l(\tau)$

$\Leftarrow$: Consider the case where $\sigma$ is $\widehat{(q,i,j)}$-near to $\tau$. Let $\mu_\sigma = \alpha \cup v$ with $v \in (\sigma \backslash \tau)$ and $\mu_\tau = \alpha \cup w$ with $w \in (\tau \backslash \sigma)$. We choose $k$ as the index of $v$ in $\mu_\sigma$ and $l$ as the index of $w$ in $\mu_\tau$. Then, $\sigma$ is $(q,k,l)$-near to $\tau$, because $d_k(\mu_\sigma) = d_l(\mu_\tau) = \alpha$. ◀

### 2.3.2 Differences

The following two statements are directly relevant to understand the Bottom-Up and Hybrid approach. As stated in Remark 12, these nice properties only hold for $(q,i,j)$-nearness. Thus it follows immediately that both notions cannot be equal. Further insight on how and why they differ may be gained by studying Theorem 13.

▶ **Lemma 10** (Upward Closure). *Let $\mu_\sigma \hookrightarrow \sigma$ and $\mu_\tau \hookrightarrow \tau$ be two $(q,i,j)$-near simplices in $\Sigma_{q+1}$. Then $\sigma$ is also $(q,i,j)$-near to $\tau$.*

**Proof.** This follows directly from the definition of $(q,i,j)$-nearness. Since $\mu_\sigma, \mu_\tau \in \Sigma_{q+1}$, they share the face $\alpha = d_i(\mu_\sigma) = d_j(\mu_\tau)$. The $(q,i,j)$-nearness follows from $\mu_\sigma \hookrightarrow \sigma$ and $\mu_\tau \hookrightarrow \tau$. ◀

Note that, by definition, the statement holds in reverse as well. Thus:

▶ **Corollary 11.** *A simplex $\sigma \in \Sigma_{>q}$ is $(q,i,j)$-near to any simplex $\tau$ iff at least one $(q+1)$-simplex $\mu_\sigma \hookrightarrow \sigma$ is $(q,i,j)$-near to $\tau$.*

▶ **Remark 12.** Lemma 10 does not hold for $\widehat{(q,i,j)}$-nearness. Consider the subset of a flag complex in Figure 3. The red edges are contained in the $(2,1,3)$-near-digraph, but not in the $\widehat{(2,1,3)}$-near-digraph. According to Lemma 10, if $\sigma$ is $(2,1,3)$-near to $\tau$, then the same should hold for $\sigma'$ and $\tau'$. This is not the case, because $d_1(\sigma') = (0234)$ and $d_3(\tau') = (0135)$ do not share a $(2)$-face.

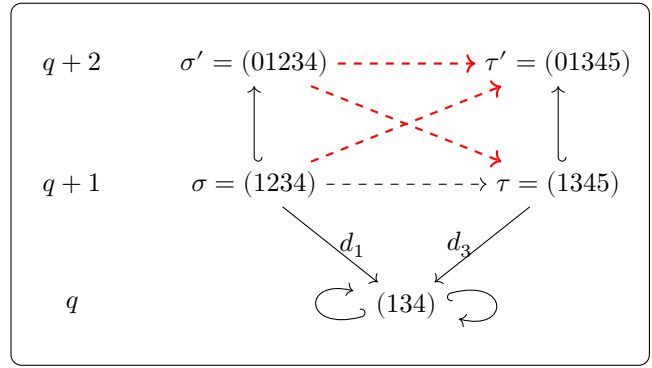

**Figure 3** Subset of some flag complex that demonstrates that Lemma 10 does not hold for $\widehat{(q,i,j)}$-nearness.

The major difference between the two definitions is the role of the indices $i$ and $j$. When considering the $\widehat{(q,i,j)}$-nearness of $\sigma$ and $\tau$, indices $i$ and $j$ are used to refer to $\sigma_i$ and $\tau_j$ respectively. In the new definition, $i$ and $j$ are indices for subsimplices $\mu_\sigma, \mu_\tau \in \Sigma_{q+1}$, where $\mu_{\sigma_i} = \sigma_{\tilde{i}}$. The index $\tilde{i}$ then falls somewhere in the range of $[i, \dim(\sigma) - q + i - 1]$. This interval originates from the fact that $i$ vertices of $\mu_\sigma$ occur before $\sigma_{\tilde{i}}$ and the other $q + 1 - i$ vertices of $\mu_\sigma$ after $\sigma_{\tilde{i}}$.

▶ **Theorem 13.** *Let $\sigma \in \Sigma_n$ and $\tau \in \Sigma_m$. The simplices $\sigma$ and $\tau$ are $(q,i,j)$-near iff there exists $i' \geq i$, which decomposes $\sigma$ into $(\sigma_\triangleleft, \sigma_{i'}, \sigma_\triangleright)$ and $j' \geq j$, which decomposes $\tau$ into $(\tau_\triangleleft, \tau_{j'}, \tau_\triangleright)$ such that*
- *$\sigma_\triangleleft$ and $\tau$ share an $(i-1)$-face $\hat{\sigma}_\triangleleft$,*
- *$\sigma_\triangleright$ and $\tau$ share a $(q-i)$-face $\hat{\sigma}_\triangleright$,*

- $\tau_\triangleleft$ and $\sigma$ share a $(j-1)$-face $\hat{\tau}_\triangleleft$,
- $\tau_\triangleright$ and $\sigma$ share a $(q-j)$-face $\hat{\tau}_\triangleright$ and
- $(\hat{\sigma}_\triangleleft, \hat{\sigma}_\triangleright) = (\hat{\tau}_\triangleleft, \hat{\tau}_\triangleright)$.

**Proof.** See Appendix B. ◀

## 3 Algorithms to Compute $(q,i,j)$-Digraphs

We describe four algorithms for calculating $\hat{\mathcal{Q}}$ respectively $\mathcal{Q}$ given the flag complex $\Sigma$ of a simple graph $G$. All algorithms require $\Sigma_{\geq q}, q, i, j$ as parameters, assuming the flag complex $\Sigma$ was previously computed via software like `flagser` [4, 5]. As $\Sigma_{\geq q}$ represents the vertex set of the output graph, the algorithmic task tackled is to find the output edge set $E^Q$ (i.e. all pairs of $(q,i,j)$-near simplices). While, by definition, $\mathcal{Q}$ includes all possible self-loops, we omit them as to not unnecessarily blow up the result. Pseudocode for all the algorithms presented here is given in Appendix A.

### 3.1 Top-Down Approach

First we present a formalization of the algorithmic ideas of the code (which was provided upon request) used to conduct the experiments described in [8]. It is thus an algorithm to compute $\hat{\mathcal{Q}}$, to which we refer to as GET_$\hat{\mathcal{Q}}$_TD. We extend it to compute $\mathcal{Q}$ as well and refer to that algorithm as GET_$\mathcal{Q}$_TD.

The main idea is straightforward: Iterate through all pairs of simplices $\sigma, \tau \in \Sigma_{\geq q}$ and check for $\widehat{(q,i,j)}$-nearness respectively $(q,i,j)$-nearness. Both checks are sketched in Figure 4.

**Checking $\widehat{(q,i,j)}$-nearness of $\sigma, \tau \in \Sigma_{\geq q}$**

Given $\sigma, \tau \in \Sigma_{\geq q}$ we first check if they even share a $q$-sized face as it forms a necessary condition. If so, we check if one embeds into the other (criterion [I]), or if faces$(d_i(\sigma), q) \cap$ faces$(d_j(\tau), q)$ is nonempty, as this would provide a simplex $\alpha \in \Sigma_q$ which fulfills criterion [II].

**Checking $(q,i,j)$-nearness of $\sigma, \tau \in \Sigma_{\geq q}$**

Again, we first check if they even share a $q$-sized face, and one is included in the other (criterion [I]). For criterion [II] we first compute $\alpha_\sigma := \{d_i(\mu_\sigma) \mid \mu_\sigma \in \text{faces}(\sigma, q+1)\}$ and analogously $\alpha_\tau := \{d_j(\mu_\tau) \mid \mu_\tau \in \text{faces}(\tau, q+1)\}$ and then check if $\alpha_\sigma \cap \alpha_\tau$ is nonempty in order to find an $\alpha$ with the desired attributes.

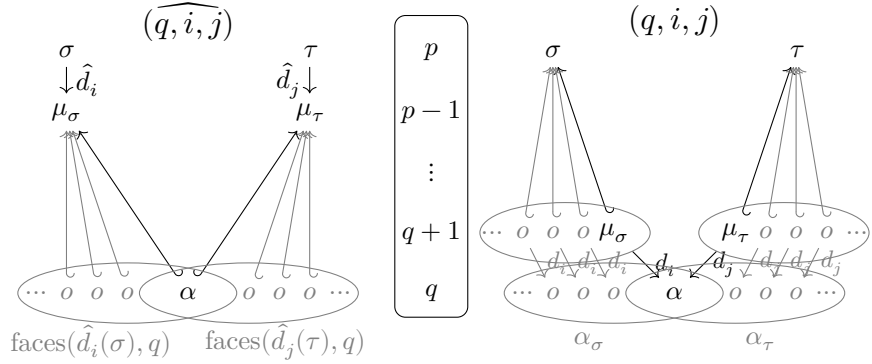

**Figure 4** The Top-Down algorithm checking for fulfillment of criterion [II] for both the original (left) and the novel definition (right).

It is straightforward to check that both algorithms work correctly and thus, by iterating over all $\sigma, \tau \in \Sigma_{\geq q}$ one correctly computes $\hat{\mathcal{Q}}$ respectively $\mathcal{Q}$.

## 3.2 Hybrid Approach

We present the new algorithm used in [11] for computing $E^{\mathcal{Q}}$ which utilizes the hybrid approach. We refer to it as GET_$\mathcal{Q}$_HYBRID.

We partition the output edge set $E^{\mathcal{Q}}$ into three (not necessarily disjoint) sets $E^{\mathcal{Q}} = E_{\hookrightarrow} \cup E_{q+1} \cup E_{\twoheadrightarrow}$, which we compute in that order. See Figure 5 for a comic-strip style example.

1. The inclusion edges $E_{\hookrightarrow}$ are the ordered pairs $(\sigma, \tau) \in \Sigma_{\geq q} \times \Sigma_{\geq q}$ such that $\sigma \hookrightarrow \tau$ (i.e. which fulfill condition [I]). We compute $E_{\hookrightarrow}$ in a top-down pass and cache the $i,j$-cofaces of all $\alpha \in \Sigma_q$ and $\delta^+_{\mathcal{Q}_{\hookrightarrow}}(\mu)$ for every $\mu \in \Sigma_{q+1}$.
2. $E_{q+1}$ contains all pairs of simplices in $\Sigma_{q+1}$ which are $(q,i,j)$-near by condition [II]. The computation starts at the (necessary) shared $q$-simplex $\alpha$ and uses the previously cached cofaces.
3. Combining $E_{q+1}$ and $\delta^+_{\mathcal{Q}_{\hookrightarrow}}(\mu)$ using the property of upward closure we find $E_{\twoheadrightarrow}$, the set of all simplex pairs that are $(q,i,j)$-near by condition [II].

### Criterion [I]: Computing $E_{\hookrightarrow}$.

Once the flag complex is known, we calculate all inclusion pairs $\sigma \hookrightarrow \tau$ in an inverse manner: Instead of starting at $\sigma$ and finding all supersimplices that include it (expensive!), we start at $\tau$ and instead enumerate all its faces. Iterating over all $\tau$ results in all inclusions being enumerated:

▶ **Lemma 14.** *Let $E_{\hookrightarrow}$ be the subset of edges in $\mathcal{Q}$ due to criterium [I]. Then:*

$$E_{\hookrightarrow} = \bigcup_{\tau \in \Sigma_{>q}} \left( \{ \bigcup_{d=q}^{\dim(\tau)} \text{faces}(\tau, d) \} \times \tau \right). \tag{1}$$

**Proof.** The incoming edges in the inclusion graph connect any simplex $\tau$ to all its faces: $\delta^-_{\mathcal{Q}_{\hookrightarrow}}(\tau) = \bigcup_{d=q}^{\dim(\tau)} \text{faces}(\tau, d)$. Because $E = \bigcup_{y \in V}(\delta^-(x) \times x)$ for *any* graph $(V, E)$, the result follows. ◀

Computing $E_{\hookrightarrow}$ is thus achieved by nested `for` loops iterating over the simplices $\tau$ and then (by dimension) all their faces. Enumerating faces is a simple process, as we merely have to remove vertices from $\tau$. Each face $\sigma$ discovered is then saved in an edge list forming $E_{\hookrightarrow}$. Furthermore, to allow for $\mathcal{O}(1)$ access later, all $(q+1)$-dimensional cofaces of all $q$-simplices are cached in a separate lookup-table. Note that the asymptotic runtime is linear in the number of inclusion edges found.

### Criterion [II]: Computing $E_{q+1}$.

With all cases of $(q, i, j)$-nearness by criteria [I] being enumerated with above algorithm we turn our attention to criteria [II], starting at the first interesting layer of the flag complex, $\Sigma_{q+1}$.

▶ **Lemma 15.** *Let $E_{q+1} \subseteq E_{\rightarrow}$ be the subset of all edges between $(q+1)$-simplices. Then*

$$E_{q+1} = \bigcup_{\alpha \in \Sigma_q} \Big( \mathrm{coF}_i(\alpha) \times \mathrm{coF}_j(\alpha) \Big). \tag{2}$$

**Proof.** Let $(\sigma, \tau) \in E_{q+1}$. Because self-loops are not part of $\mathcal{Q}$ and they are both of the same dimension, $\sigma \neq \tau$ and they can only be $(q, i, j)$-near by condition [II]. In this case $\mu_\sigma = \sigma$ and $\mu_\tau = \tau$ and by definition, there exists a $q$-simplex $\alpha$ for which $d_i(\mu_\sigma) = \alpha = d_j(\mu_\tau)$. As $\mathrm{coF}_i$ is the left inverse of $d_i$, this is equivalent to the existence of a $q$-simplex $\alpha$ such that $\mathrm{coF}_i(\alpha) = \mu_\sigma$ and $\mathrm{coF}_j(\alpha) = \mu_\tau$ and thus "$\subseteq$" holds. The other direction is clear by definition. ◀

Again, this immediately implies a suitable algorithm of three nested `for`-loops, looping over $\alpha \in \Sigma_q$ and then all $\sigma \in \mathrm{coF}_i(\alpha)$ respectively $\tau \in \mathrm{coF}_j(\alpha)$.

### Criterion [II]: Full $E_{\rightarrow}$.

With $E_{\hookrightarrow}$ and $E_{q+1}$ computed in the previous steps, we now exploit the upward closure property: For every pair $(\mu_s, \mu_\tau) \in E_{q+1}$, all combinations of their super-simplices are also contained in $E_{\rightarrow}$.

▶ **Lemma 16.** *Let $E_{\rightarrow}$ denote the set of edges in $\mathcal{Q}$ that are due to criterion [II]. Then:*

$$E_{\rightarrow} = \{\delta^+_{\mathcal{Q}_{\hookrightarrow}}(\mu_\sigma) \times \delta^+_{\mathcal{Q}_{\hookrightarrow}}(\mu_\tau) \mid (\mu_\sigma, \mu_\tau) \in E_{q+1}\}. \tag{3}$$

**Proof.** Lemma 10 immediately implies "$\supseteq$" and "$\subseteq$" follows from Corollary 11. ◀

Crucially, both $\mathcal{Q}_{\hookrightarrow}$ and $E_{q+1}$ are already known at this point in the algorithm. The sets $\delta^+_{\mathcal{Q}_{\hookrightarrow}}(\sigma)$ can be cached during the inclusion step for all $\sigma \in \Sigma_{q+1}$ with minimal overhead or otherwise found using simple graph search.

### Hybrid Algorithm for $\widehat{(q, i, j)}$-nearness.

The hybrid approach may also be adapted to accelerate the computation of $\hat{\mathcal{Q}}$. In that case, we cache $coF_i(\mu)$ and $coF_j(\mu)$ for all $\mu \in \Sigma_{>q}$ instead of just $\Sigma_{q+1}$ during the top-down pass. This is necessary as $\mu_\sigma$ and $\mu_\tau$ may be of arbitrary dimension for $\widehat{(q, i, j)}$-nearness, as opposed to fixed at $q+1$ in $(q, i, j)$-nearness. Additionally, during the upwards pass, the order of iteration between inclusions and cofaces is reversed.

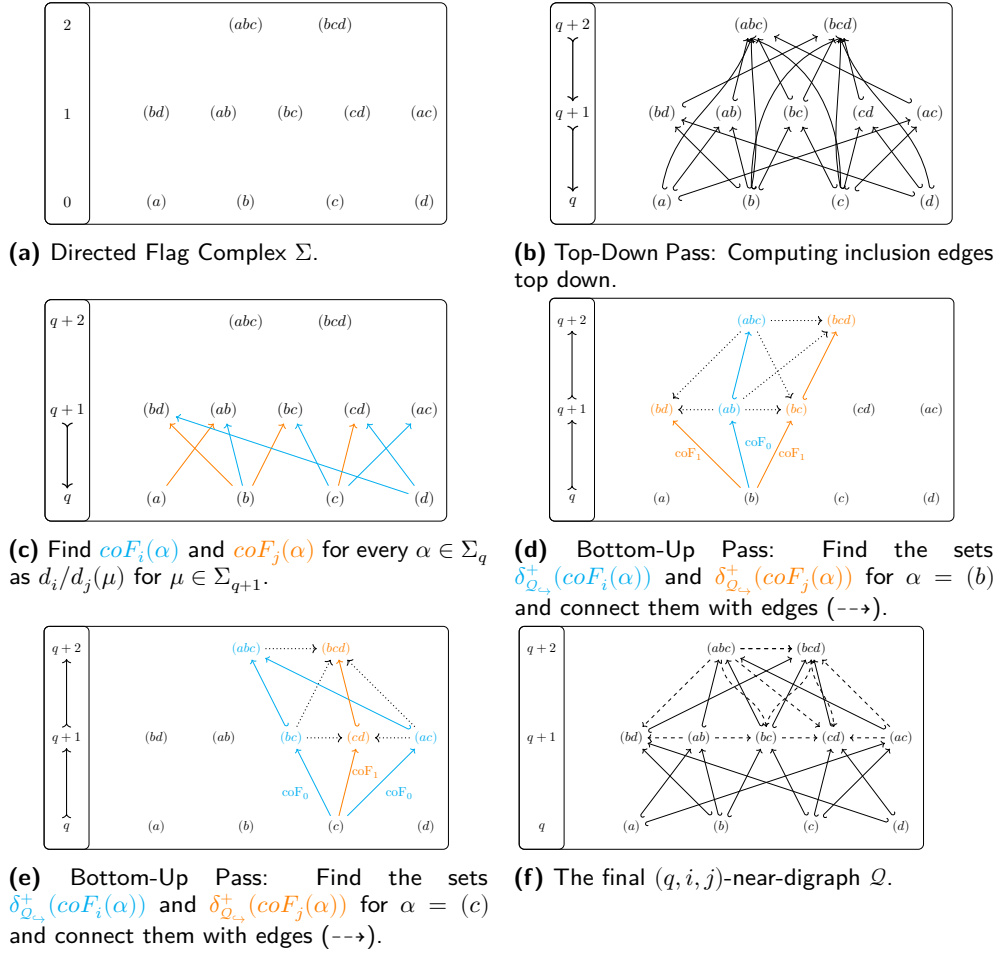

**Figure 5** Step-by-step computation of the $(q, i, j)$-near-digraph of the graph from Figure 1.

## 4 Computational Time Complexity

▶ **Proposition 17.** *The asymptotic runtime of the Top-Down algorithms* GET_$\hat{\mathcal{Q}}$_TD *and* GET_$\mathcal{Q}$_TD *is bounded by* $\Omega(|\Sigma_{\geq q}|^2)$.

**Proof.** $\Omega(1)$ is certainly a lower bound for a single check of $\widehat{(q, i, j)}$-nearness respectively $(q, i, j)$-nearness between two simplices, even if accounted for potential optimisation. Thus, to check for all potential combinations of $\sigma$ and $\tau$, we require at least $\Omega(|\Sigma_{\geq q}|^2)$ effort. ◀

▶ **Proposition 18.** *The runtime of the Hybrid algorithm* GET_$\mathcal{Q}$_HYBRID *is asymptotically bound by*
$$\mathcal{O}\left(|\mathcal{Q}| \cdot (D+1)^{q+3}\right),$$
*where $D$ is the maximal simplex dimension in $\Sigma$.*

**Proof.** The time complexity of computing $E_{\hookrightarrow}$ and $E_{q+1}$ is linear in the number of edges found. Propagating these edges to higher dimensions introduces a subtle problem though: edges may be found multiple times. Simplices $\sigma$ and $\tau$ can share up to $\binom{D+1}{q+1}$ $q$-faces $\alpha$.

Each of those shared faces $\alpha$ may be contained in $D-q$ different $(q+1)$-faces $\mu_\sigma$ of $\sigma$ (and analogously for $\mu_\tau$ and $\tau$): The face $\mu_\sigma$ can be written as $(\alpha_0...\alpha_{i-1}\sigma_k\alpha_{i+1}...\alpha_{q+1})$, where $\sigma_k$ may be any of the up to $D$ vertices of $\sigma$ not yet preset in $\alpha$. As such, $(D-q)^2$ is an upper bound for the number of $(q+1)$-face pairs $\mu_\sigma \hookrightarrow \sigma, \mu_\tau \hookrightarrow \tau$ that share $\alpha$. Multiplying with the number of possible $\alpha$ yields $\binom{D+1}{q+1}(D-q)^2 \in \mathcal{O}((D+1)^{q+3})$, and multiplying it further with the number of edges yields what is to prove. ◀

While this upper bound is only legitimate in the worst-case scenario of dense graphs, it appears that the potential number of occurrences of each edge in the algorithm scales polynomially in $D$ and exponentially in $q$. In all practical applications this becomes less of a concern, as graphs investigated by $q$-analysis are typically far from dense which reduces the number of duplicated edges drastically. In order to remove duplicate edges efficiently, the final output edge list is stored as a Hash Set.

In practice, $q$ is a constant chosen by the user and typically not in the double digits. In [11] they furthermore recommend artificially clipping the flag complex to $\Sigma_{\leq D_{\max}}$ with $q < D_{\max} \leq 2q$ to avoid severe interpretability problems and to enhance the results. Thus, for practical purposes, both $q$ and $D$ can be considered constants, meaning every edge is found a constant number of times. With this assumption, the algorithm becomes time-optimal:

▶ **Theorem 19.** *Consider $q$ a constant and let $D$ be restrained by the constant $D_{max}$. Then the Hybrid algorithm* GET_Q_HYBRID *has an asymptotic runtime of $\mathcal{O}(\mathcal{Q})$ and is thus asymptotically time optimal.*

To show the difference in practice, we include the benchmark table from [11]. We can observe a speedup of $\sim 3.7 \times 10^5$ for the $(4,i,j)$-near-digraph of BBP at $q=4$ for the hybrid algorithm. As expected, this can be attributed to the sparseness of the output graph, which has a density of around $\sim 8.2 \times 10^{-5}$.

|  |  |  |  | Top-Down | | Hybrid | |
|---|---|---|---|---|---|---|---|
| Graph | Nodes | Edges | $q$ | $\widehat{(q,i,j)}$ | $(q,i,j)$ | $\widehat{(q,i,j)}$ | $(q,i,j)$ |
| *C. Elegans* [9] | 279 | 2194 | 3 | 10.68s | 11.16s | 27.93ms | 25.72ms |
| Erdős–Rényi | 1000 | 50k | 3 | 16.98s | 17.34s | 765.64μs | 455.97μs |
| BBP [6] | 31k | 7.6M | 4 | 2062.22s | 2054.21s | 14.60ms | 5.57ms |
| BBP [6] | 31k | 7.6M | 3 | $> 48$h | $> 48$h | 45.85s | 36.18s |

■ **Table 1** Benchmark results for the `Rust` implementation of the Hybrid and Top-Down algorithm.

## 5  Parallelization

Each individual step of the hybrid algorithm (computing $E_\hookrightarrow$, $E_{q+1}$ and $E_\rightarrow$) by itself is embarrassingly parallel. Taking the upward pass as an example, it is trivial to distribute the elements of $\Sigma_q$ uniformly onto arbitrary many threads and task them with computing the cartesian product of their $i/j$-cofaces and their supersimplices. However, merging the output of this step (or any other before, as their complete result is necessary for computing $E_\rightarrow$) introduces a serious bottleneck, as we can't easily pre-allocate memory for the results, because the output size/location of each individual thread is a-priori unknown. We report on three different approaches of dealing with this problem:

**Mutual Exclusion** One possible option is using mutex locks to manage the access of each thread to the shared datastructure containing the result. This leads to small performance improvements for low number of threads, especially when writes to the result are done in batches, but for high parallelization this can actually hurt performance due to high contention.

**Split-and-Merge** Instead of sharing a datastructure, the results of each thread are merged after each step using a divide-and-conquer-approach. Merging simple datastructures such as maps or lists is linear in time and due to divide-and-conquer every element in the final result is merged only $log_2(T)$ times, where $T$ is the number of total threads. Simultaneously, the time it takes to find the $n$ output results is reduced by a factor of $T$. Assuming we can also parallelize the merging operations, this approach has a theoretical runtime of $\mathcal{O}(\frac{n}{T} + \frac{n \cdot log_2(T))}{T})$, as opposed to $\mathcal{O}(n)$ for single-thread. In practice, this approach does accelerate the overall computation, but memory bandwidth leads to diminishing returns for increasing thread counts.

**Bottom-Up** The motivation for computing the inclusions and cofaces in a top-down manner is that finding the right vertices to insert (finding cofaces bottom-up) is computationally harder than finding faces (removing vertices). Harder is not impossible though: With a slow-down-factor of $|V|$ it is possible to compute the set $coF_i(\sigma)$ directly by checking for every vertex $v$ in the input graph whether $(\sigma_0 \sigma_1 ... \sigma_{i-1} v ... \sigma_n)$ is also a directed simplex in the input graph $G$. This way the precomputations (Step 1 and 2 of the Hybrid approach) can be skipped entirely: $coF_i / coF_j(\alpha)$ can be computed directly and $\delta^+_{Q_\rightarrow}(coF_i / coF_j(\alpha))$ can be computed by finding cofaces recursively. While this leads to significantly higher computational effort, it means each thread can produce output edges of $Q$ entirely independently.

**Benchmarks** Overall, parallelizing the Hybrid algorithm has led to limited success, with diminishing returns for adding more cores, as can be seen in Table 2. While the bottom-up approach significantly increases computational effort, it scales much better with increased parallelization, making it potentially viable in cluster-computing environments.

| Graph | $q$ | cores | Single-thread | Mutex | Split-and-Merge | Bottom-Up |
|---|---|---|---|---|---|---|
| BBP | 4 | 2 | 640ms | 540ms | 540ms | 1201s |
| BBP | 4 | 8 | 640ms | 490ms | 470ms | 300s |
| BBP | 4 | 32 | 690ms | 610ms | 450ms | 81s |

**Table 2** Benchmarks for different parallelization methods on the Blue Brain Project graph [6] for different number of cores running on an AMD EPYC 7543.

## 6 Discussion

Overall, the Hybrid approach and its algorithms are a significant step in making directed $q$-analysis useful and practical, as explored in [11]. However, even if Theorem 19 might imply that the Hybrid approach is the last approach one ever requires, there are, unfortunately, still some drawbacks. The most significant one is the memory usage of the whole method:

**Space Complexity** Depending on the choice of $q$ and input graph, both the size of the output $(q, i, j)$-near-digraph or even the intermediate flag complex $\Sigma$ can significantly exceed the size of the input. As both are simultaneously stored in RAM, the algorithm is more hindered by memory than compute, as for big inputs the size of $\mathcal{Q}$ can exceed RAM capacity within minutes of computation. In order to compute massive $(q, i, j)$-digraphs, the result would need to be dynamically written to disk. This represents a problem for the Hybrid approach, as cofaces and inclusions need to be cached in RAM for the upwards propagation of $E_{q+1}$. In contrast, the true Bottom-Up algorithm only needs to keep $\delta_{\mathcal{Q}_\to}(\mu)$ for some $\mu \in \Sigma$ in RAM. With a subsequent deduplication step, it might be the only feasible option for truly massive input graphs.

**Future Work**  As hinted at before, the Bottom-Up approach is unavoidable in some scenarios. Luckily, there is also untapped potential for its development. Its significant slow-down compared to the Hybrid approach mostly comes from the necessity to compute cofaces again and again in the worst possible way — scanning all vertices of $V$ to see what fits. We see potential in investigating, for the specific scenario in question, the sweet-spot of time/space-trade-offs, which most certainly lies in between the extremes presented here.

## References

[1]  Ronald H. Atkin. "An algebra for patterns on a complex. I." In: *International Journal of Man-Machine Studies* 6 (1974), pp. 285–307. ISSN: 0020-7373. DOI: 10.1016/S0020-7373(74)80024-6.

[2]  Ronald H. Atkin. "An algebra for patterns on a complex. II." In: *International Journal of Man-Machine Studies* 8.5 (1976), pp. 483–498. ISSN: 0020-7373. DOI: 10.1016/s0020-7373(76)80015-6.

[3]  Ronald H. Atkin. "The Methodology of Q-Analysis Applied to Social Systems." In: *Systems Methodology in Social Science Research: Recent Developments*. Ed. by Roger Cavallo. Dordrecht: Springer Netherlands, 1982, pp. 45–74. ISBN: 978-94-017-3204-8. DOI: 10.1007/978-94-017-3204-8_4. URL: https://doi.org/10.1007/978-94-017-3204-8_4.

[4]  Daniel Lütgehetmann. *flagser*. https://github.com/luetge/flagser. 2017-2021.

[5]  Daniel Lütgehetmann et al. "Computing persistent homology of directed flag complexes." In: *Algorithms (Basel)* 13.1 (2020), Paper No. 19, 18. DOI: 10.3390/a13010019.

[6]  Henry Markram et al. "Reconstruction and simulation of neocortical microcircuitry." In: *Cell* 163.2 (2015), pp. 456–492.

[7]  Alessandro Motta et al. "Dense connectomic reconstruction in layer 4 of the somatosensory cortex." In: *Science* 366.6469 (Nov. 2019). DOI: 10.1126/science.aay3134.

[8]  Henri Riihimäki. "Simplicial $q$-Connectivity of Directed Graphs with Applications to Network Analysis." In: *SIAM Journal on Mathematics of Data Science* 5.3 (2023), pp. 800–828. DOI: 10.1137/22M1480021. eprint: https://doi.org/10.1137/22M1480021. URL: https://doi.org/10.1137/22M1480021.

[9]  Lav R Varshney et al. "Structural properties of the Caenorhabditis elegans neuronal network." In: *PLoS Comput Biol* 7.2 (2011), e1001066.

[10]  Felix Windisch. "A Novel Definition of Directed q-Nearness: Comparative Analysis and Algorithms." Available at https://doi.org/10.3217/z7ys0-gcj46. MA thesis. Technical University of Graz, 2024.

[11]  Felix Windisch and Florian Unger. *Fast Directed q-Analysis for Brain Graphs*. accepted April 2025, currently in production. arXiv: 2501.04596 [q-bio.QM]. URL: https://arxiv.org/abs/2501.04596.

## A  Pseudocode

We present pseudocode for both Top-Down and Bottom-Up algorithms.

### A.1  Top-Down Algorithm

**Algorithm 1** Naive algorithm.

---
1: **function** GET_$\mathcal{Q}$_NAIVE($\Sigma$, q, i, j)
2:     $\mathcal{Q} \leftarrow$ unconnected_graph($\Sigma_{\geq q}$)
3:     **for all** $(\sigma, \tau)$ **in** $(\Sigma_{\geq q} \times \Sigma_{\geq q})$ **do**                                     $\triangleright$ $[\mathcal{O}(\Sigma_{>q}^2)]$
4:         **if** IS_Q_NEAR($\sigma, \tau$, q, i, j)) **then**
5:             $\mathcal{Q}$.add($\sigma \rightarrow \tau$)                                                           $\triangleright$ Add edge
6:     **return** $\mathcal{Q}$
---

**Algorithm 2** Check $\widehat{(q,i,j)}$-nearness of two simplices.

---
1: **function** IS_$\hat{Q}$_NEAR($\sigma, \tau$,q,i,j)
2:     **if** $|\sigma \cap \tau| > q$ **then**
3:         **return** $(|\operatorname{faces}(d_i(\sigma), q) \cap \operatorname{faces}(d_j(\tau), q)| > 0 \ \ \textbf{or}\ \ \sigma \hookrightarrow \tau)$
4:     **return** false
---

**Algorithm 3** Check $(q, i, j)$-nearness of two simplices.

---
1: **function** IS_Q_NEAR($\sigma, \tau$,q,i,j)
2:     **if** $|\sigma \cap \tau| > q$ **then**
3:         $\alpha_\sigma \leftarrow [\operatorname{faces}(\sigma, q+1).\mathtt{map}(x \rightarrow \partial_i(x))]$
4:         $\alpha_\tau \leftarrow [\operatorname{faces}(\tau, q+1).\mathtt{map}(x \rightarrow \partial_j(x))]$
5:         **return** $(|\alpha_\sigma \cap \alpha_\tau| > 0 \ \ \textbf{or}\ \ \sigma \hookrightarrow \tau)$
6:     **return** false
---

### A.2  Hybrid Algorithm

**Algorithm 4** Hybrid Algorithm for computing $\mathcal{Q}$.

---
1: **function** GET_$\mathcal{Q}$($\Sigma$)
2:     $E_\hookrightarrow \leftarrow$ GET_$E_\hookrightarrow$($\Sigma$)                                                               $\triangleright\ \Theta(|E_\hookrightarrow|)$
3:     $\mathtt{coF}_i, \mathtt{coF}_j \leftarrow$ GET_SIMPLEX_COFACES($\Sigma_{q+1}$)        $\triangleright\ \Theta(|\Sigma_{q+1}|)$
4:     $E_\rightarrow \leftarrow$ GET_$E_\rightarrow$($\Sigma, \mathtt{coF}_i, \mathtt{coF}_j$)                                $\triangleright\ \Theta(|E_\rightarrow|)$
5:     **return** $(\Sigma, E_\hookrightarrow \cup E_\rightarrow)$                                                        $\triangleright\ \Theta(|\mathcal{Q}|)$
---

**▉ Algorithm 5** Method for computing the inclusion graph.

---
1: **function** GET_$E_{\hookrightarrow}(\Sigma)$
2:    $E_{\hookrightarrow} \leftarrow \emptyset$
3:    **for** $p$ **from** $q+1$ **to** D **do**
4:       **for** $\sigma \in \Sigma_p$ **do**
5:          **for** d **from** q **to** $\dim(\sigma)-1$ **do**     ▷ $\dim(\sigma)-1$ to avoid self-loops
6:             **for** $\tau \in \text{faces}(\sigma, d)$ **do**
7:                $E_{\hookrightarrow}.\text{add}(\tau \hookrightarrow \sigma)$
8:    **return** $E_{\hookrightarrow}$
---

**▉ Algorithm 6** Construct the coface data structure.

---
1: **function** GET_SIMPLEX_COFACES($\Sigma_{q+1}$)
2:    $\text{coF}_i = \textbf{new Map}\ (\Sigma_q \to \text{List} < \Sigma_{q+1} >)$
3:    $\text{coF}_j = \textbf{new Map}\ (\Sigma_q \to \text{List} < \Sigma_{q+1} >)$
4:    **for** $\sigma \in \Sigma_{q+1}$ **do**
5:       $\text{coF}_i[d_i(\sigma)].\text{append}(\sigma)$
6:       $\text{coF}_j[d_j(\sigma)].\text{append}(\sigma)$
7:    **return** $(\text{coF}_i, \text{coF}_j)$
---

**▉ Algorithm 7** Compute $E_{q+1}$ using coface maps.

---
1: **function** GET_$E_{q+1}(\Sigma_q, \text{coF}_i, \text{coF}_j)$
2:    $E_{q+1} \leftarrow \emptyset$
3:    **for** $\alpha \in \Sigma_q$ **do**
4:       **for** $\sigma \in \text{coF}_i(\alpha)$ **do**
5:          **for** $\tau \in \text{coF}_j(\alpha)$ **do**
6:             $E_{q+1}.\text{add}(\sigma \to \tau)$
7:    **return** $E_{q+1}$
---

**Algorithm 8** Compute $(q, i, j)$-nearness using coface maps and inclusion.

1: **function** GET_$E_\rightarrow$($\Sigma$, coF$_i$, coF$_j$)
2: $\quad E_\rightarrow \leftarrow$ **new** HashSet<Edge>
3: $\quad$ **for** $\alpha \in \Sigma_q$ **do**
4: $\quad\quad$ **for** $\mu_\sigma \in$ coF$_i(\alpha)$ **do**
5: $\quad\quad\quad$ **for** $\mu_\tau \in$ coF$_j(\alpha)$ **do**
6: $\quad\quad\quad\quad$ **for** $\sigma \in \delta^+_{Q_\hookrightarrow}(\mu_\sigma)$ **do**
7: $\quad\quad\quad\quad\quad$ **for** $\tau \in \delta^+_{Q_\hookrightarrow}(\mu_\tau)$ **do**
8: $\quad\quad\quad\quad\quad\quad$ $E_\rightarrow$.add($\sigma \rightarrow \tau$)
9: $\quad$ **return** $E_\rightarrow$

**Algorithm 9** Hybrid Algorithm for computing $\widehat{\mathcal{Q}}$

1: **function** GET_$\widehat{\mathcal{Q}}$($\Sigma$)
2: $\quad \widehat{E}_\hookrightarrow \leftarrow$ GET_$E_\hookrightarrow$($\Sigma$) $\quad\quad\quad\quad\quad\quad\quad\quad\quad\quad\quad\quad\quad\quad\quad\quad\triangleright \Theta(|E_\hookrightarrow|)$
3: $\quad$ coF$_i$, coF$_j \leftarrow$ GET_SIMPLEX_COFACES($\Sigma_{\geq q}, \Sigma_{\geq q}$) $\quad\triangleright \Theta(|\Sigma_{\geq q}|)$
4: $\quad \widehat{E}_\rightarrow \leftarrow$ GET_$\widehat{E}_\rightarrow$($\Sigma$) $\quad\quad\quad\quad\quad\quad\quad\quad\quad\quad\quad\quad\quad\quad\quad\triangleright \Theta(|\widehat{E}_\rightarrow|)$
5: $\quad$ **return** $(\Sigma, \widehat{E}_\hookrightarrow \cup \widehat{E}_\rightarrow)$ $\quad\quad\quad\quad\quad\quad\quad\quad\quad\quad\quad\quad\triangleright \Theta(|\widehat{\mathcal{Q}}|)$

**Algorithm 10** Compute $\widehat{(q, i, j)}$-near relations using the inclusion graph

1: **function** GET_$\widehat{E}_\rightarrow$($\Sigma$, coF$_i$, coF$_j$)
2: $\quad E_\rightarrow \leftarrow \emptyset$
3: $\quad$ **for** $\alpha \in \Sigma_q$ **do**
4: $\quad\quad$ **for** $\mu_\sigma \in \delta^+_{Q_\hookrightarrow}(\alpha)$ **do**
5: $\quad\quad\quad$ **for** $\mu_\tau \in \delta^+_{Q_\hookrightarrow}(\alpha)$ **do**
6: $\quad\quad\quad\quad$ **for** $\sigma \in$ coF$_i(\mu_\sigma)$ **do**
7: $\quad\quad\quad\quad\quad$ **for** $\tau \in$ coF$_j(\mu_\tau)$ **do**
8: $\quad\quad\quad\quad\quad\quad$ $\widehat{E}_\rightarrow$.add($\sigma \rightarrow \tau$)
9: $\quad$ **return** $\widehat{E}_\rightarrow$

## B  Additional Proofs

The major difference between the two definitions is the role of the indices $i$ and $j$. When considering the $\widehat{(q, i, j)}$-nearness of $\sigma$ and $\tau$, indices $i$ and $j$ will be used to refer to $\sigma_i$ and $\tau_j$ respectively. In the new definition, $i$ and $j$ are indices for subsimplices $\mu_\sigma, \mu_\tau \in \Sigma_{q+1}$, where $\mu_{\sigma_i} = \sigma_{\tilde{i}}$. The index $\tilde{i}$ then falls somewhere in the range of $[i, \dim(\sigma) - q + i - 1]$. This interval originates from the fact that $i$ vertices of $\mu_\sigma$ occur before $\sigma_{\tilde{i}}$ and the other $q + 1 - i$ vertices of $\mu_\sigma$ after $\sigma_{\tilde{i}}$.

At first glance, it would appear that this relaxation would lead to a less strict definition, but there is another hidden assumption at play.

▶ **Theorem 20.** *Let $\sigma \in \Sigma_n$ and $\tau \in \Sigma_m$. The simplices $\sigma$ and $\tau$ are $(q, i, j)$-near iff there exists $i' \geq i$, which decomposes $\sigma$ into $(\sigma_\triangleleft, \sigma_{i'}, \sigma_\triangleright)$ and $j' \geq j$, which decomposes $\tau$ into $(\tau_\triangleleft, \tau_{j'}, \tau_\triangleright)$ such that*

- $\sigma_\triangleleft$ and $\tau$ share an $(i-1)$-face $\hat{\sigma}_\triangleleft$,
- $\sigma_\triangleright$ and $\tau$ share a $(q-i)$-face $\hat{\sigma}_\triangleright$,
- $\tau_\triangleleft$ and $\sigma$ share a $(j-1)$-face $\hat{\tau}_\triangleleft$,
- $\tau_\triangleright$ and $\sigma$ share a $(q-j)$-face $\hat{\tau}_\triangleright$ and
- $(\hat{\sigma}_\triangleleft, \hat{\sigma}_\triangleright) = (\hat{\tau}_\triangleleft, \hat{\tau}_\triangleright)$.

**Proof.** We show both directions separately. $\Leftarrow$: Assume there exist indices $i', j'$ as described in the theorem for some simplices $\sigma, \tau$. Then a shared face $\alpha$ can be constructed as $(\sigma_\triangleleft \cap \tau) \cup (\sigma_\triangleright \cap \tau)$. The first intersection will contain $i$ vertices and the second $(q-i+1)$ vertices due to the shared $(i-1)$- and $(q-i)$-face respectively. Thus, $\dim(\alpha) = (i+(q-i+1))-1 = q$. Then $\alpha = d_i((\alpha_0, ..., \alpha_{i-1}, \sigma_{i'}, \alpha_i, ..., \alpha_q)) = d_i(\mu_\sigma) \hookrightarrow \sigma$. The same construction to find $\mu_\tau$ completes this direction and the last condition guarantees that $d_i(\mu_\sigma) = d_j(\mu_\tau)$.

$\Rightarrow$: Let $\sigma$ and $\tau$ be $(q,i,j)$-near. By definition, there exist $(q+1)$-simplices $\mu_\sigma \hookrightarrow \sigma, \mu_\tau \hookrightarrow \tau$ such that $d_i(\mu_\sigma) = \alpha = d_j(\mu_\tau)$ for some $\alpha \in \Sigma_q$. Choose $i'$ as the lowest possible value such that $\mu_\triangleleft = (\mu_{\sigma_0}\mu_{\sigma_1}...\mu_{\sigma_{i-1}})$ is still contained in $\sigma_\triangleleft$. By definition of $(q,i,j)$-nearness, $\mu_\triangleleft \hookrightarrow \alpha$, and since $\alpha \hookrightarrow \tau$ and $\alpha$ contains $i$ vertices of $\sigma_\triangleleft$, the simplex $\sigma_\triangleleft$ shares a $(i-1)$-face with $\tau$. Now $\sigma_\triangleright$ must contain all vertices of $\mu_\triangleright = (\mu_{\sigma_{i+1}}\mu_{\sigma_{i+2}}...\mu_{\sigma_n})$, which are in turn contained in $\alpha$ and thus also in $\tau$, making $\mu_\triangleright$ a shared $(q-i)$-face between $\sigma_\triangleright$ and $\tau$. The other conditions follow by a symmetric argument for $\tau$. ◀